\documentclass[aps,prl,twocolumn,groupedaddress]{revtex4-1}
\usepackage{graphicx}
\usepackage{amsmath,amsfonts,mathtools,amssymb}
\usepackage[colorlinks=true,linkcolor=blue,citecolor=red,bookmarksnumbered=true]{hyperref}
\usepackage{mathpazo}
\usepackage{times}
\usepackage{color}

\makeatletter
\def\overl@ss#1#2{\vcenter{\offinterlineskip
        \ialign{$\m@th#1\hfil##\hfil$\crcr#2\crcr<\crcr } }}
\def\overgr@at#1#2{\vcenter{\offinterlineskip
        \ialign{$\m@th#1\hfil##\hfil$\crcr#2\crcr>\crcr } }}
\def\gl{\mathrel{\mathpalette\overl@ss>}}
\def\lg{\mathrel{\mathpalette\overgr@at<}}
\makeatother

\def\diag{\mathop{\rm diag}\nolimits}

\def\Real{\mathbb{R}}

\def\Re{\mathop{\rm Re}\nolimits}
\def\Im{\mathop{\rm Im}\nolimits}

\def\re{\mathrm{re}}
\def\im{\mathrm{im}}

\def\@#1{{\mathbf{#1}}}
\def\_#1{{\mathrm{#1}}}

\def\step{\mathrm{step}}
\def\zbg{\mathrm{zbg}}
\def\nzbg{\mathrm{nzbg}}
\def\osnzbg{{1\mathrm{snzbg}}}

\def\R{\mathbb{R}}

\def\be{\begin{equation}}
\def\ee{\end{equation}}
\def\bse{\begin{subequations}}
\def\ese{\end{subequations}}
\def\eqref#1{(\ref{#1})}

\long\def\paragraph#1{\par\smallskip\textbf{#1.}}

\allowdisplaybreaks

\def\reftitle#1{``#1''}

\begin{document}
\title{Nonlinear interactions between solitons and dispersive shocks in focusing media}
\author{Gino Biondini$^{1,2}$}
\author{Jonathan Lottes$^1$}
\affiliation{$^1$Department of Mathematics, State University of New York at Buffalo, Buffalo, New York 14260, USA\\
    $^2$Department of Physics, State University of New York at Buffalo, Buffalo, New York 14260, USA
    }
\date{This manuscript was compiled on \today}

\begin{abstract}
Nonlinear interactions in focusing media between traveling solitons and the dispersive shocks produced by an initial discontinuity 
are studied using the one-dimensional nonlinear Schr\"odinger equation.
It is shown that, when solitons travel from a region with nonzero background towards a region with zero background,
they always pass through the shock structure without generating dispersive radiation.
However, their properties (such as amplitude, velocity and shape)
change in the process.  
A similar effect arises when solitons travel from a region with zero background towards a region with nonzero background,
except that, depending on its initial velocity, in this case the soliton may remain trapped inside the shock-like structure indefinitely.
In all cases, the new soliton properties can be determined analytically.
The results are validated by comparison with numerical simulations.
\end{abstract}
\maketitle




\medskip

\paragraph{Introduction}
%
A common way to study the response of a nonlinear system is to consider Riemann problems, i.e.,
the evolution of a jump discontinuity between two uniform values of the initial datum.
In dispersive nonlinear media, Riemann problems can give rise to dispersive shock waves (DSWs),
which are non-stationary coherent wave structures
and arise in many different physical contexts, including water waves,
the atmosphere, optics and Bose-Einstein condensates.
As a result, considerable effort has been devoted to the study of DSW formation, propagation and interactions 
\cite{Whitham1976,Kamchatnov,GP1974,elgurevich,PLA1995v203p77,SJAM59p2162,BK2006,HA2007}.
The topic has also attracted renewed interest in recent years \cite{BEHM2016,EH2016,prl120p144101,sirev59p3,jmp50p091406,pre97p032218}.

An ubiquitous tool in nonlinear physics is the nonlinear Schr\"odinger (NLS) equation, 
which is a universal model for the evolution of the envelope of weakly nonlinear dispersive wave trains \cite{JMP46p133}.
The NLS equation arises in a wide variety of physical settings, including
deep water waves, fiber optics, plasmas and Bose-Einstein condensates \cite{AS1981,Agrawal2007,IR2000,PS2003}.
The NLS equation is also a completely integrable infinite-dimensional Hamiltonian system \cite{AS1981,NMPZ1984,FT1987,APT2004}. 
This means that the initial value problem can be solved by the inverse scattering transform (IST) \cite{ZS1972,ZS1973}.

Typically, DSWs are produced either in the small dispersion, or semiclassical, limit, which arises when the dispersive effects are small compared
to nonlinear ones, or in the long-time asymptotics.
For the focusing NLS equation, the asymptotic behavior of solutions in the semiclassical limit with zero background (ZBG) has been
studied extensively, with sech-shaped input~\cite{millerkamvissis1998,bronskikutz1999,KMM2003}, 
generalizations thereof~\cite{TVZ2004,lyng2012}, 
and box-like input~\cite{JM2013,EKT2016,TE2016}.
In all these cases, the focusing nonlinearity results in the formation of highly peaked oscillations in a localized region of space.

The situation is more complicated with nonzero background (NZBG), due to the presence of modulational instability (MI),
namely, the fact that a constant background is unstable to long-wavelength perturbations \cite{ZO2009}. 
A quantitative description of the nonlinear stage of MI 
for generic localized perturbations of a constant background was recently obtained in~\cite{BM2016,PRE94p060201R,BM2017}. 
The corresponding behavior, which is comprised of two quiescent states separated by a central wedge with modulated periodic oscillations,
was later found to arise in a broad class of NLS-type evolution equations describing a variety of focusing nonlinear media in~\cite{SIREV2018}, 
and was recently observed experimentally in~\cite{arxiv2018elrandouxsuret}.
Related scenarios arise from Riemann problems.
Special cases of Riemann problems for the focusing NLS equation were studied in \cite{bikbaev,BV2007,BKS2011}, 
and more general Riemann problems were recently considered in \cite{B2018}.
The expanding oscillatory wedge between two uniform states 
can be viewed as a DSW in focusing media.

Solitons play no role in the above discussion.
On the other hand, the focusing NLS equation admits a large variety of soliton solutions, both with ZBG \cite{ZS1972} and NZBG \cite{BK2014}.
The combined presence of solitons and dispersive radiation in focusing media with NZBG was recently shown to produce
novel phenomena such as soliton transmission, trapping and wake~\cite{BLM2018}.
The purpose of this work is to investigate nonlinear interactions arising when the solution
contains all three of the above components, namely: 
NZBG, solitons and dispersive shocks.
In particular, we study a practical scenario, namely,
the interaction between a soliton and the oscillatory wedge formed by a discontinuity in the initial condition.
We show that, when traveling from a region with nonzero background towards a region with zero background,
solitons always pass through these shock structures and retain their identity, without generating dispersive radiation.
Importantly, however, we also show that, even though the discrete eigenvalue in the scattering problem is of course time-independent,
all of the properties of the corresponding soliton (including amplitude, velocity and shape) change 
once they move from a region with NZBG to one with ZBG.

A similar scenario arises when solitons travel from a region with ZBG to one with NZBG,
except that now, depending its velocity, the soliton can remain trapped inside the shock-like structure indefinitely.
In both cases, the new soliton properties are analytically determined by computing the long-time asymptotics of solutions, and 
the results are validated by comparison with extensive numerical simulations.

\paragraph{NLS equation, ZBG and NZBG, solitons}
%
The cubic one-dimensional NLS equation is the partial differential equation
\vspace*{-1ex}
\be
\label{e:nls}
i q_t + q_{xx} + 2(|q|^2 -q_o^2) q = 0\,.
\ee
Subscripts $x$ and $t$ denote partial differentiation, and $q(x,t)$
typically describes the complex-valued envelope of oscillations.
The non-negative real parameter $q_o$ corresponds to a background amplitude \cite{gauge}.

The IST makes crucial use of the existence of a Lax pair, namely the fact that Eq.~\eqref{e:nls} is the compatibility condition $\phi_{xt} = \phi_{tx}$ of the overdetermined linear system
\vspace*{-1ex}
\be
\phi_x = X\phi\,,\qquad 
\phi_t = T\phi\,,
\label{e:LP}
\ee
with 
$X = -ik\sigma_3 + Q$ and 
$T = -i(2k^2+q_o^2-|q|^2-Q_x)\,\sigma_3 -2kQ$, where 
$\sigma_3 = \diag(1,-1)$ is the third Pauli matrix,  and
\vspace*{-1ex}
\be
Q(x,t) = \begin{pmatrix} 0 &q \\ - q^* & 0 \end{pmatrix}.
\ee
The first half of the Lax pair~\eqref{e:LP} and $q(x,t)$ are referred to as the scattering problem and the potential, respectively,
while $k$ is the scattering parameter or eigenvalue of the scattering problem.

Here we study the behavior of solutions of Eq.~\eqref{e:nls} with the following step-like boundary conditions:%
\vspace*{-0.8ex}
\begin{gather}
q(x,t) \to q_\pm\,,\qquad x\to\pm\infty,
\label{e:BCs}
\end{gather}
with $q_- = 0$, and 
where without loss of generality we can take $q_+ = q_o$ thanks to the phase invariance of the NLS equation.
We refer to Eqs.~\eqref{e:BCs} as the case of one-sided nonzero background (1SNZBG).
When $q_o =0$, they reduce to the case of ZBG \cite{ZS1972,AS1981,NMPZ1984,APT2004}.
Conversely, when $q_-\ne0$, 
one has a problem with a two-sided NZBG.

The IST in the case with ZBG was developed in the seminal work by Zakharov and Shabat~\cite{ZS1972}.
The symmetric NZBG case $|q_-| = |q_+|= q_o$ was developed in \cite{BK2014},
and was extended to the fully asymmetric case $|q_\pm|\ne0$ and $|q_-|\ne|q_+|$ in \cite{JMP55p101505},
while the one-sided NZBG was studied in \cite{PV2015}.
Recall that the IST works by associating to $q(x,t)$ time-independent scattering data via the scattering problem.
Once the scattering data are obtained from the initial conditions (ICs), the solution of Eq.~\eqref{e:nls} is then reconstructed by inverting the scattering
transform.
The scattering data are computed in terms of the Jost eigenfunctions, which are the solutions $\phi_\pm(x,t,k)$ of the Lax pair~\eqref{e:LP} that reduce to plane waves as $x\to\pm\infty$, and are therefore the nonlinearization of the Fourier modes.
The set of all complex values of $k$ for which the Jost eigenfunctions are defined comprises the continuous spectrum $\Sigma$ of the scattering problem.
For potentials on ZBG, the continuous spectrum is simply the real $k$-axis; i.e., $\Sigma_\zbg = \Real$ \cite{ZS1972}.
For potentials on NZBG or 1SNZBG, however, the continuous spectrum acquires a subset of the imaginary axis, 
namely $\Sigma_\nzbg = \Sigma_\osnzbg = \R \cup i[-q_o,q_o]$ \cite{BK2014,PV2015}.
Moreover, for potentials on NZBG, the nonlinear analogue of the Fourier wavenumber is given by 
$\lambda = (q_o^2 + k^2)^{1/2}$.

\begin{figure}[t!]
\centering{\includegraphics[width=.45\textwidth]{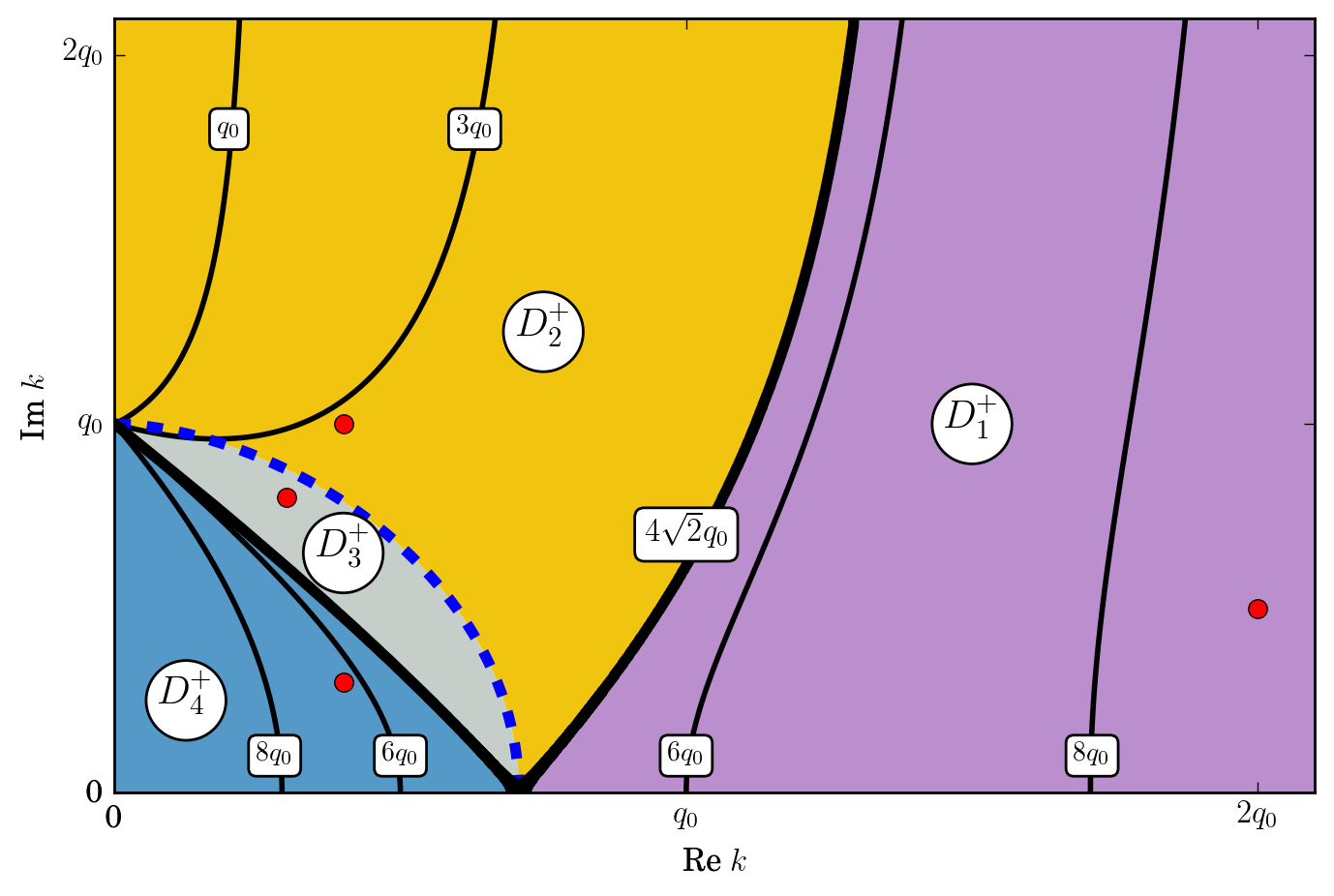}}
\caption{Contour lines of constant soliton velocity in the spectral plane for solitons on NZBG and the domains $\smash{D_1^+},\smash{D_2^+},\smash{D_3^+},\smash{D_4^+}$ resulting in the various 
interaction outcomes for a right-moving soliton (see text for details). 
The blue dashed curve is determined by a system of modulation equations (see text for details). 
Red dots: the discrete eigenvalues that generate the solitons in Fig.~\ref{f:rightmoving}.}
\label{f:Vnzbg}
\end{figure}

The discrete spectrum of the scattering problem, when present, gives rise to soliton solutions. 
In particular, each discrete eigenvalue contributes one soliton to the solution. 
Both with ZBG and with NZBG, 
all the properties of the soliton are determined explicitly by the location of the discrete eigenvalue in the complex plane. 
These properties however differ with ZBG versus NZBG.
In particular, a soliton on ZBG generated by a discrete eigenvalue $k=k_\re + i k_\im$ travels with velocity \cite{ZS1972}
\vspace*{-0.6ex}
\be
V_\zbg(k) = 4k_\re\,.
\ee
A soliton on NZBG generated by the same discrete eigenvalue, however, travels with velocity 
\vspace*{-0.6ex}
\be
V_\nzbg(k) = 2(k_\re + k_\im \lambda_\re / \lambda_\im),
\ee
where $\lambda = \lambda_\re + i \lambda_\im$.
A contour plot of constant soliton velocity in the spectral plane for solitons on NZBG is given in Fig.~\ref{f:Vnzbg}.
For solitons on ZBG, the curves of constant soliton velocity are obviously given by vertical lines, 
towards which the contour lines in Fig.~\ref{f:Vnzbg} tend asymptotically as $\Im k\to\infty$.
Note also $V_\nzbg(k)>V_\zbg(k)$ for all $k$ in the first quadrant.

\paragraph{Interactions between solitons and DSWs: Set-up}
The simplest realization of ICs consistent with the boundary conditions~\eqref{e:BCs}, is a ``pure step'' problem, namely
\vspace*{-0.4ex}
\be
q_\mathrm{step}(x,0) = q_o H(x),
\label{e:stepIC}
\ee
where $H(x)$ is the Heaviside step function, defined as $H(x) = 0$ for $x<0$ and $H(x)=1$ for $x>0$.
The above IC results in the formation of an oscillatory ``wedge'' (or DSW) in the region $0< x < 4\sqrt2 q_o t$, 
to the left of which the solution is negligible and to the right of which the solution is approximately equal to the background $q_o$.
Inside the wedge, the solution can be described as a slow modulation of the traveling wave (elliptic) solutions of the focusing NLS equation,
\cite{BKS2011} (see below for details).

Here we consider situations arising from a combination of the above step ICs and a traveling soliton of
the focusing NLS equation.
Specifically,
we consider the case of a soliton generated by a discrete eigenvalue located at $k=k_o$ and initially positioned at $x = X_o$. 
If $X_o>0$, the soliton is initially positioned to the right of the step (i.e., on a NZBG), whereas if $X_o<0$ it is initially positioned to the left of the step
(i.e., on a ZBG).
Denoting by $V_o$ the initial velocity of the soliton,
if $\Re(k_o)>0$, the soliton will travel to the right (i.e., $V_o>0$), whereas if $\Re(k_o)<0$, it will travel to the left (i.e., $V_o<0$).
Thus, if $X_oV_o>0$, the soliton will travel away from the DSW, whereas if $X_oV_o<0$ it will move towards it.
The most interesting scenario 
is obviously the latter.
We therefore consider two main cases:
\\[0.6ex]
\hglue1em 1. A left-moving soliton initially placed on NZBG (i.e., to the right of the initial discontinuity), corresponding to $V_o<0$ and $X_0>0$.
\\[0.4ex]
\hglue1em 2. A right-moving soliton initially placed on ZBG (i.e., to the left of the initial discontinuity), corresponding to $V_o>0$ and $X_0<0$.

It should be noted that, numerically, the ICs corresponding to the above situations are realized in a different way depending on whether $X_o$ is positive or negative.
If $X_o<0$, one can simply \textit{add} a one-soliton solution of the focusing NLS equation with ZBG to the pure step IC.
If $X_o>0$ instead, one should \textit{multiply} the pure step IC by the one-soliton solution of the focusing NLS equation with NZBG.
Once the appropriate ICs have been realized, the time integration of Eq.~\eqref{e:nls} was performed using an eighth-order Fourier split-step method. 
As in \cite{BM2016,PRE94p060201R,BM2017,SIREV2018}, the computed time evolution is only accurate up to the time at which 
round-off error grows to $O(1)$ \cite{accuracy}.


%
%

\begin{figure}[t!]
\smallskip
\centering{\includegraphics[width=.475\textwidth]{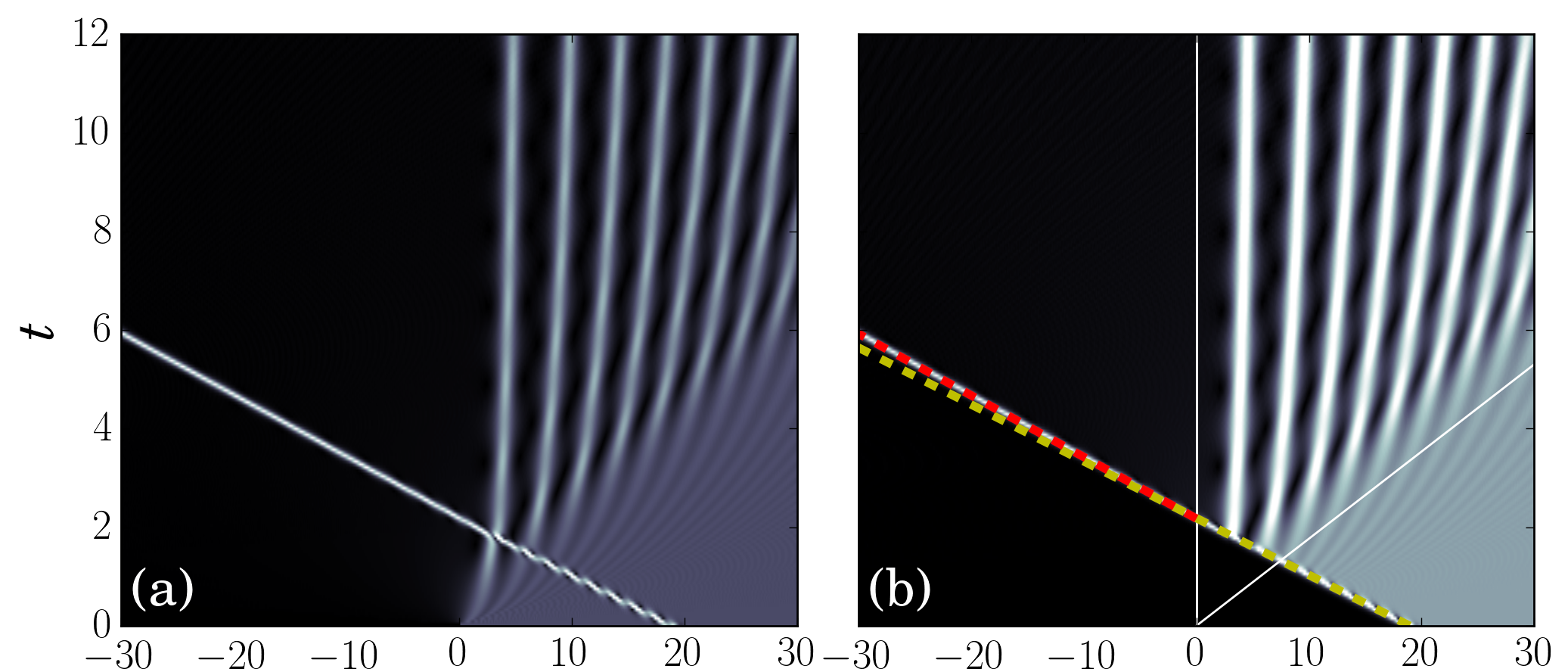}}
\centering{\includegraphics[width=.475\textwidth]{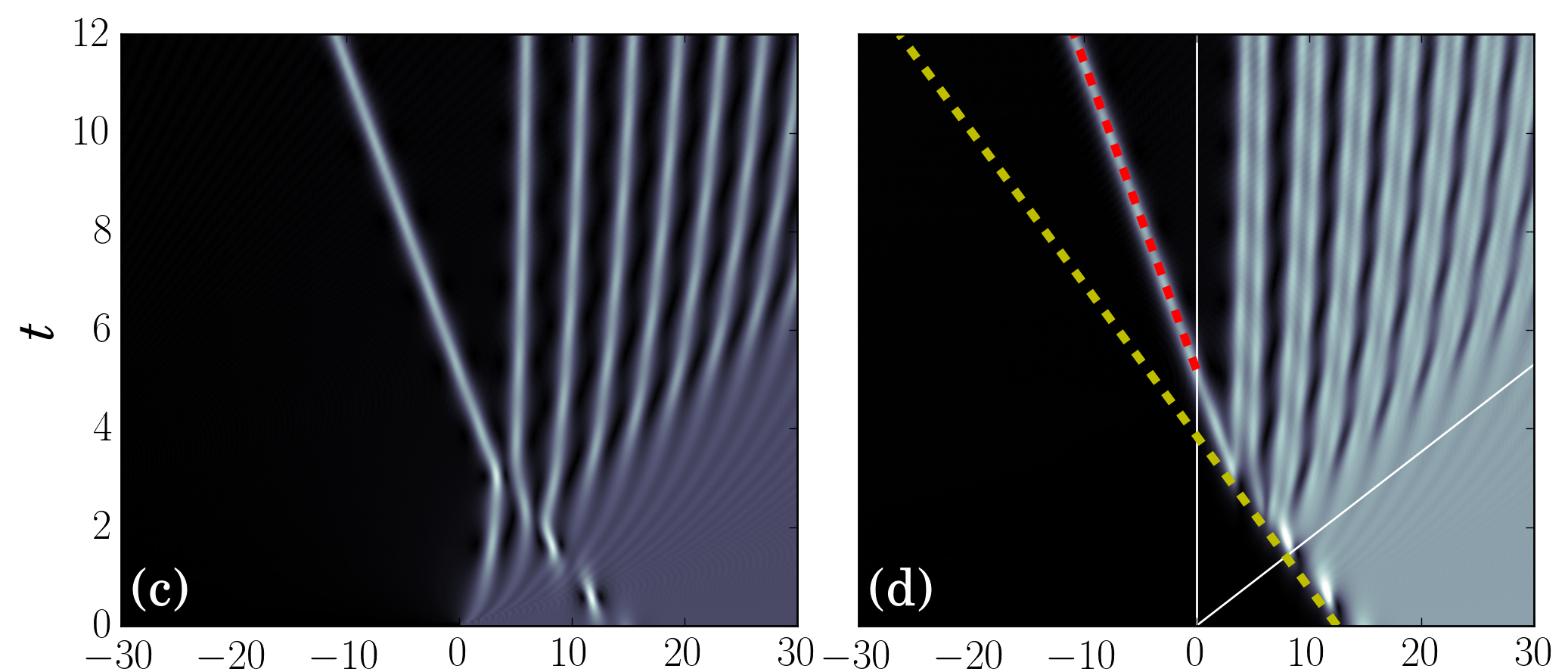}}
\centering{\includegraphics[width=.475\textwidth]{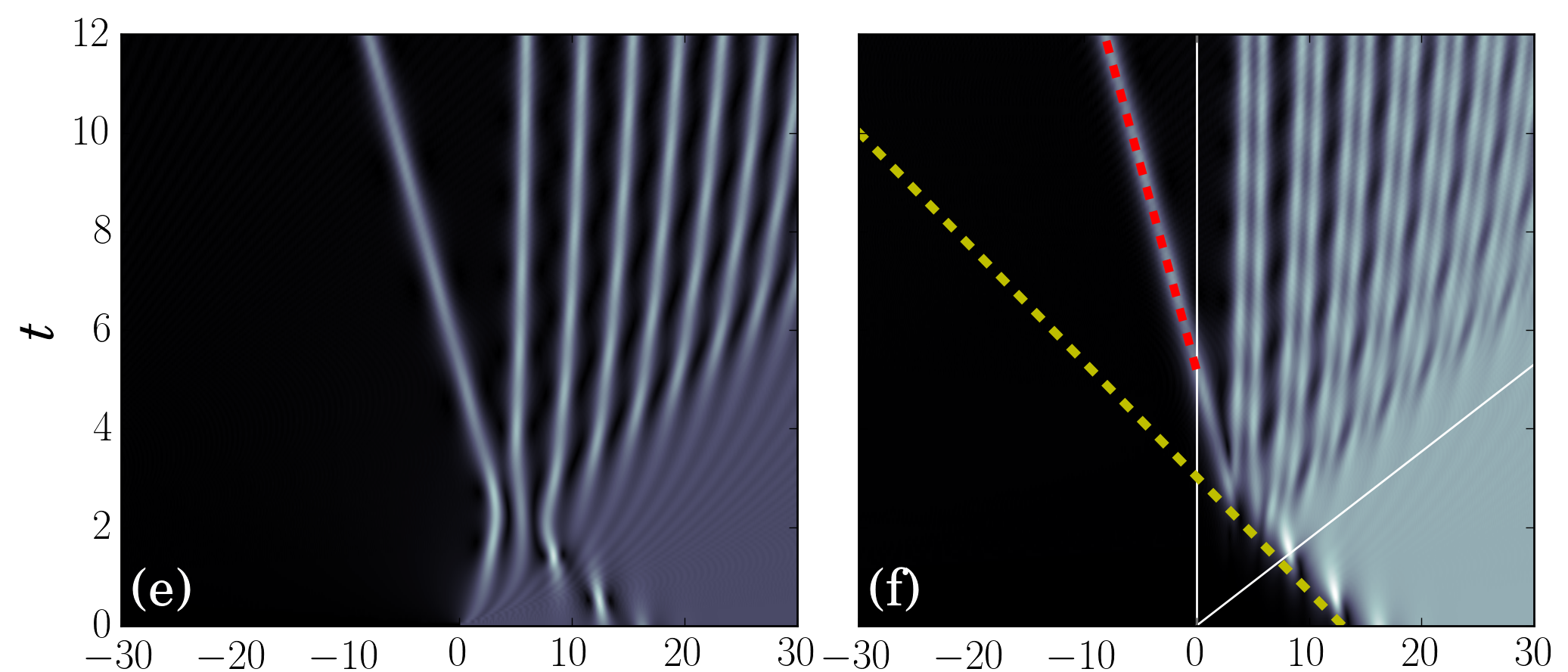}}
\centering{\includegraphics[width=.475\textwidth]{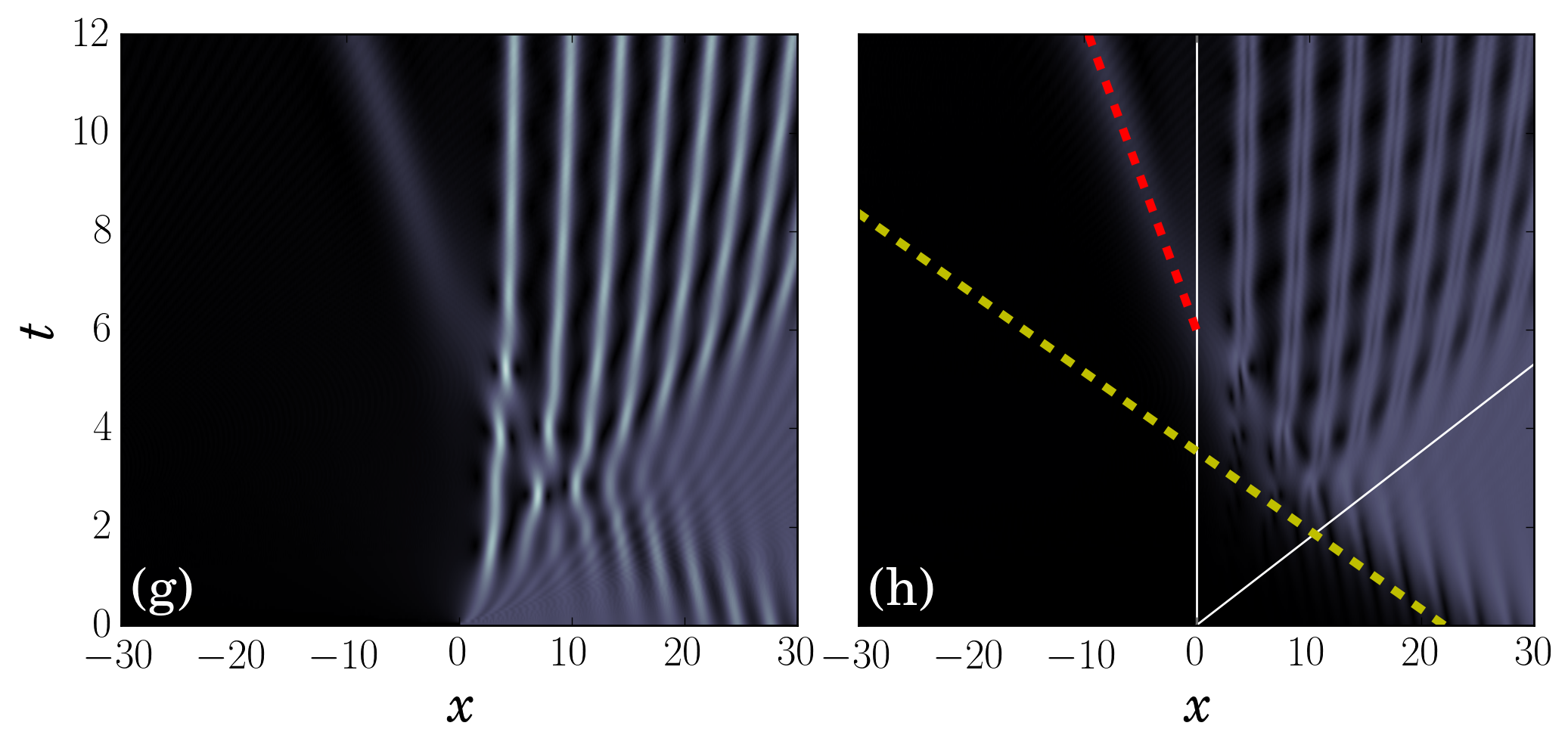}}
\caption{Solutions of the focusing NLS equation with a left-moving soliton initially placed on a NZBG $q_o=1$ and a step to ZBG at $x=0$.
Left column: Density plot of $|q(x,t)|$ as a function of $x$ and $t$.
Right column: Density plot of the difference between $q(x,t)$ and the solution $q_\step(x,t)$ produced by the pure step IC~\eqref{e:stepIC}, 
illustrating the permanent effect of the nonlinear interactions.
Solid white lines: boundaries $x=0$ and $x=4\sqrt{2}q_o t$ of the wedge. 
Dashed lines: initial trajectory (yellow, velocity $V_\nzbg$) 
and the trajectory of the soliton after it exits the wedge (red, velocity $V_\zbg$),
demonstrating the change of the soliton velocity.
Top row [(a) and~(b)]: $k_o = -2 + 1.5i \in \smash{D_1^-}$, resulting in $V_\zbg = -8$ and $V_\nzbg = -8.68$.
Second row [(c) and~(d)]: $k_o = -0.4+i \in \smash{D_2^-}$, resulting in $V_\zbg = -1.6$ and $V_\nzbg = -3.24$.
Third row [(e) and~(f)]: $k_o = -0.3+0.8i \in \smash{D_3^-}$, resulting in $V_\zbg = -1.2$ and $V_\nzbg = -4.29$.
Bottom row [(g) and~(h)]: $k_o = -0.4+0.3i \in \smash{D_4^-}$, resulting in $V_\zbg = -1.6$ and $V_\nzbg = -6.21$.
}
\label{f:leftmoving}
\kern-\medskipamount
\end{figure}

\begin{figure}[t!]
\smallskip
\centering{\includegraphics[width=.475\textwidth]{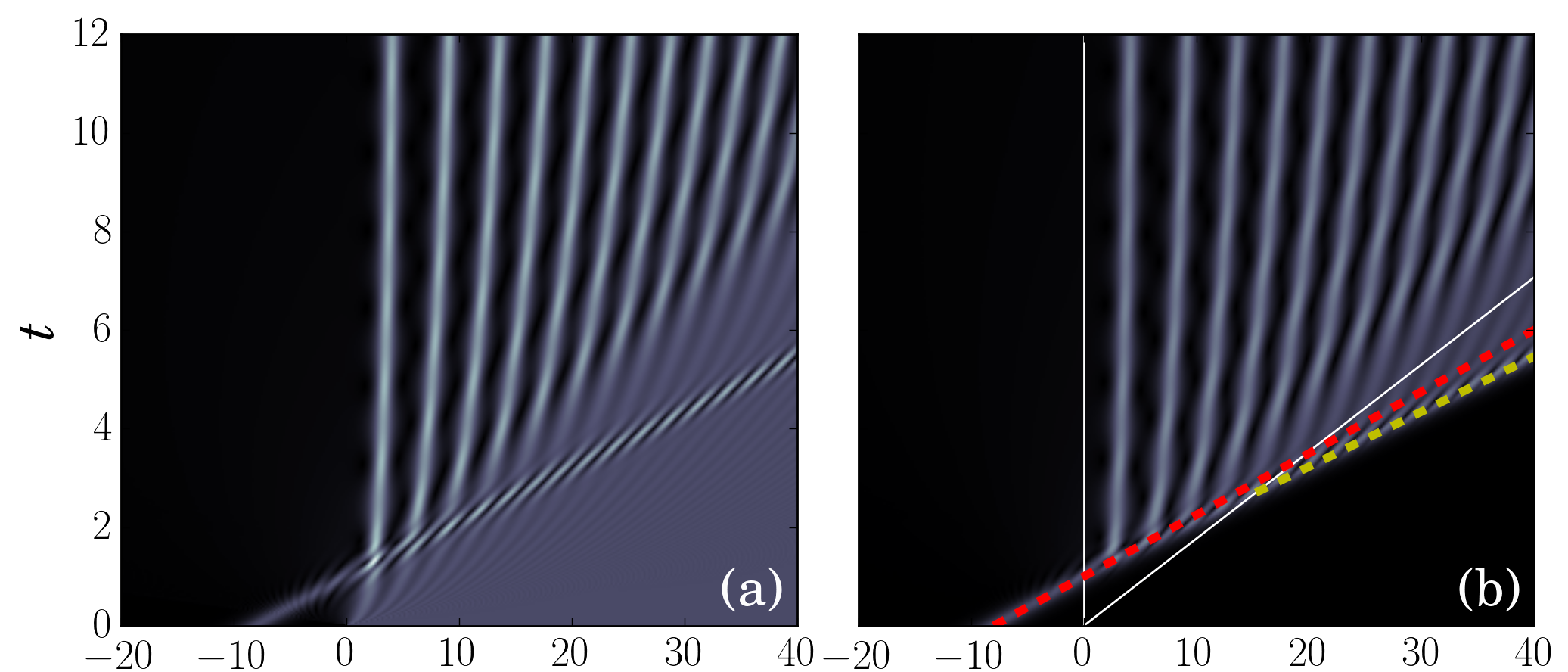}}
\centering{\includegraphics[width=.475\textwidth]{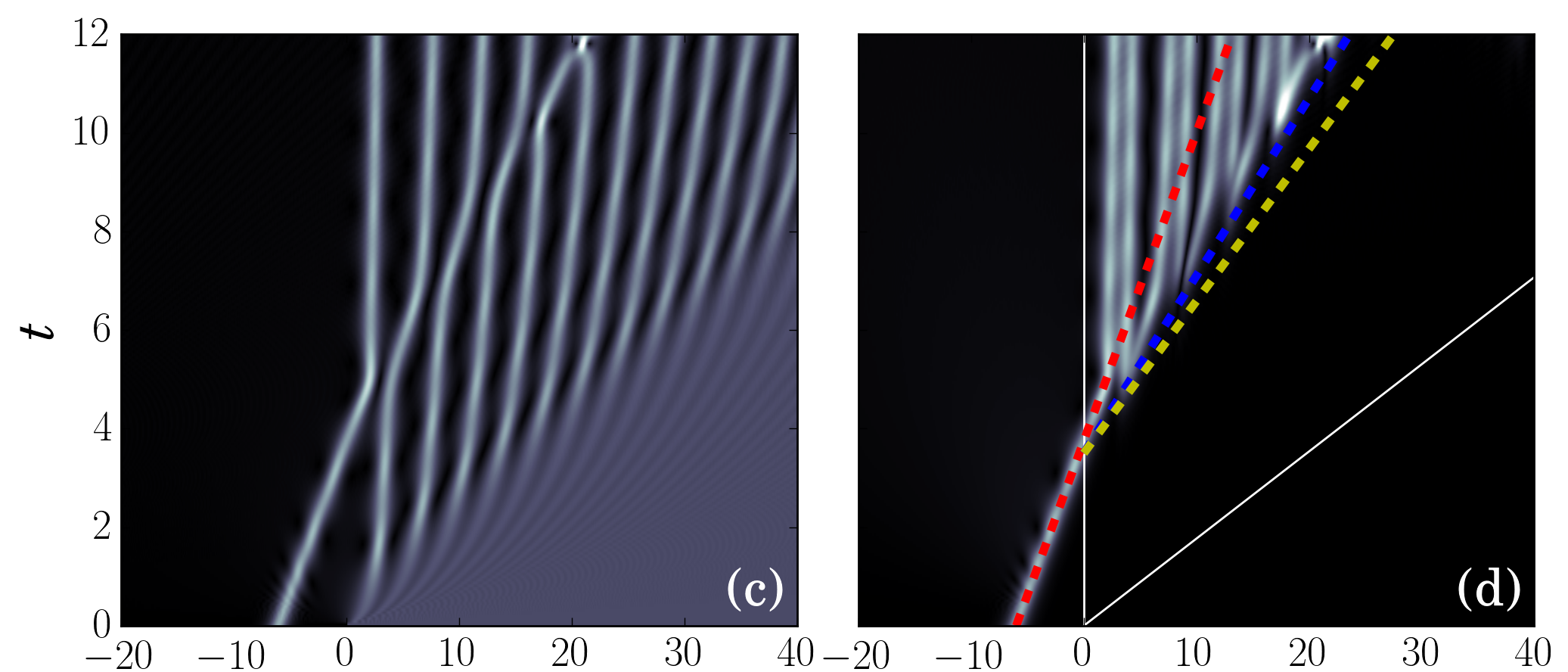}}
\centering{\includegraphics[width=.475\textwidth]{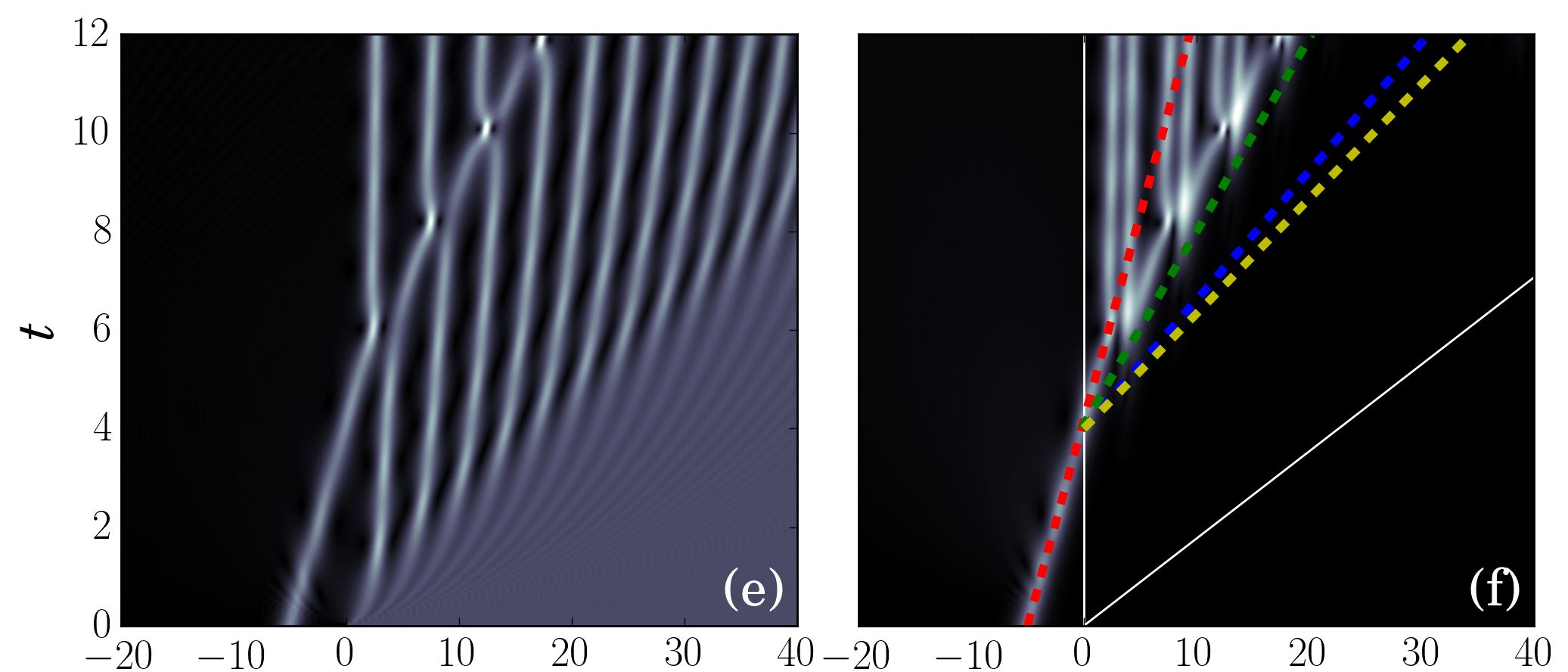}}
\centering{\includegraphics[width=.475\textwidth]{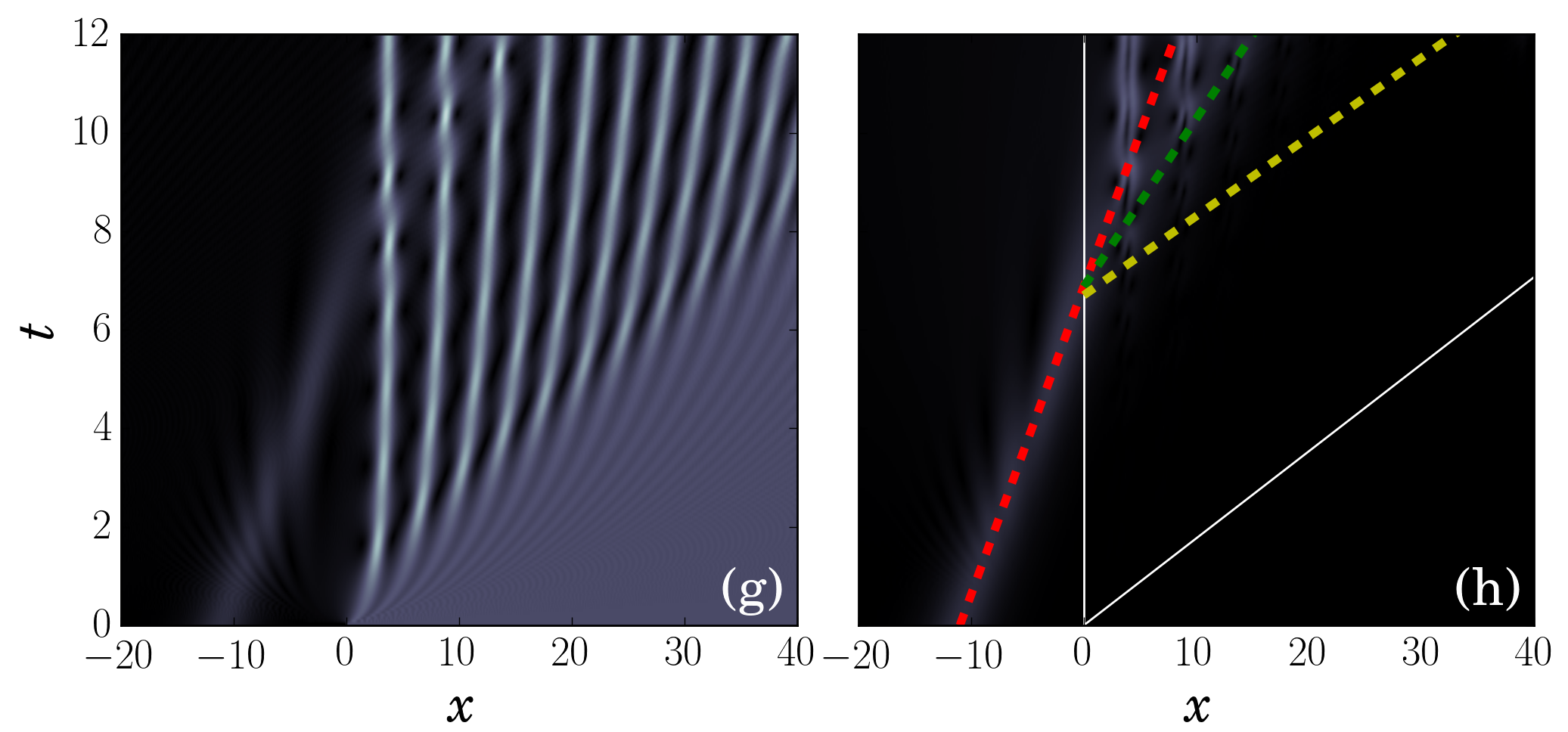}}
\caption{Same as Fig.~\ref{f:leftmoving}, but for a right-moving soliton initially placed on ZBG together with a step to the NZBG $q_o=1$ at $x=0$.
Dashed lines: initial trajectory (red, velocity $V_\zbg$) and
trajectory of the soliton after it exits the wedge (yellow, velocity $V_\nzbg$, or blue, velocity $V_s$, or green, velocity $V_\ast$, see text for details).
The discrete eigenvalues are the symmetric counterparts of those in Fig.~\ref{f:leftmoving}: namely, $k_o\mapsto - k_o^*$.
Correspondingly, all values of $V_\zbg$ and $V_\nzbg$ are the opposite of those in Fig.~\ref{f:leftmoving}.
Additionally, 
$V_s=2.78$ in~(d),
$V_s = 3.83$ and $V_\ast = 2.55$ in~(f),
and $V_\ast = 2.93$ in~(h).
}
\label{f:rightmoving}
\kern-\medskipamount
\end{figure}

\paragraph{Interaction between solitons and DSWs: Results}
Figure~\ref{f:leftmoving} presents the results of numerical simulations corresponding to case~1 above (left-moving soliton initially placed to the right of the step),
while Fig.~\ref{f:rightmoving} corresponds to case~2 (right-moving soliton initially placed to the left of the step).6
In each case, the left column shows density plots of the solution amplitude $|q(x,t)|$.
while the right column shows the difference between the solution to the left and that obtained from the pure step IC~\eqref{e:stepIC}, 
which provides a direct visual illustration of the nonlinear interaction effects.
Each row corresponds to a different choice of discrete eigenvalue.

In case~1 (soliton initially on a NZBG and traveling leftward toward the DSW)
we observe that, for all choices of discrete eigenvalue, the soliton is transmitted through the DSW and emerges as a soliton on ZBG.
Importantly, however, Fig.~\ref{f:leftmoving} clearly shows that all of the properties of the soliton 
(that is, its amplitude, width, velocity, and its breather-like versus traveling-wave nature)
are changed after it has traveled through the oscillatory structure.

The change of the soliton features may be surprising, since 
the properties of the soliton are completely determined by the location of the discrete eigenvalue, 
and both the continuous and discrete spectrum of the scattering problem are independent of time.
As we show below, however, these changes are not a numerical artifact, and are indeed reflective of the true nonlinear dynamics of the system.

As shown in Fig.~\ref{f:rightmoving},
a similar outcome arises in case~2
(soliton initially on a ZBG and traveling rightward toward the DSW).
In addition, however, here there are cases (e.g., second and third row)
in which the soliton does not escape the oscillatory wedge, and remains trapped there forever,
as is easily seen by comparing the soliton trajectory with that of the wedge boundary \cite{escape}.

\paragraph{Long-time asymptotics}
We emphasize that none of the velocities in the trajectories shown in the right column of Figs.~\ref{f:leftmoving} and~\ref{f:rightmoving} 
were determined numerically,
and all of them are determined analytically instead.
Indeed, we next show that the
numerical results of Figs.~\ref{f:leftmoving} and~\ref{f:rightmoving} can be fully characterized by studying the long-time asymptotics of solutions
of the focusing NLS equation with nonzero background.

A full calculation of the long-time asymptotics is beyond the scope of this work, 
so here we limit ourselves to presenting the essential details.
Recall that, in the inverse problem of IST, the solution of the NLS equation is obtained from that of a suitable
matrix Riemann-Hilbert problem (RHP) defined in terms of the reflection coefficient and, if present, the discrete spectrum.
A key part of the RHP is a controlling phase function appearing in the jump conditions.
In the case of ZBG, this phase function is $\theta_\zbg(x,t,k) = k(x - 2kt)$ \cite{APT2004},
whereas with NZBG, $\theta_\nzbg(x,t,k) = \lambda\,(x - 2kt)$ instead \cite{BK2014}, with $\lambda(k)$ as above.
Indeed, it is precisely by looking along directions $x=Vt$ and setting $\Im[\theta(x,t,k_o)]=0$ that one finds the soliton velocities corresponding to a discrete 
eigenvalue at $k=k_o$ with ZBG and NZBG.

Importantly, however, in the long-time asymptotics of solutions with symmetric NZBG in the presence of a small disturbance initially placed near $x=0$,
the governing phase function gets modified in the wedge $x\in(-4\sqrt2q_ot,4\sqrt2q_ot)$
\cite{BM2016,BM2017}.
For those values of $x$, $\theta(x,t,k)$ is replaced by a new phase function $h(x,t,k)$ defined in terms of certain Abelian integrals \cite{BM2017}.
It was also shown in \cite{BLM2018} that it is by setting $\Im[h(x,t,k_o)]=0$ that one determines the velocity of a soliton inside the wedge.
For the one-sided NZBG studied here, the same considerations apply in the half wedge $x\in(0,4\sqrt2q_ot)$.

\paragraph{Classification of interactions}
For a left-moving soliton starting on a NZBG, 
the plots in the right column of Fig.~\ref{f:leftmoving} clearly show that, 
for the choices of $k_o$ considered, 
the soliton velocity is given by $V_\nzbg(k_o)$ before entering the oscillation structure, 
and by $V_\zbg(k_o)$ after exiting it.

The situation is more complex for a right-moving soliton starting on a ZBG.
In this case, one observes different outcomes depending on the precise location of the discrete eigenvalue.
Consider a discrete eigenvalue $k_o$ in the first quadrant of the complex plane, 
and recall the contour plot of soliton velocity with NZBG in Fig.~\ref{f:Vnzbg}.
One can distinguish four domains:
in $\smash{D_1^+}$ (purple region) and $\smash{D_4^+}$ (blue region), one has $V_\nzbg(k_o)>4\sqrt2 q_o$, 
whereas in 
in $\smash{D_2^+}$ (yellow region) and $\smash{D_3^+}$ (gray region), one has $V_\nzbg(k_o)<4\sqrt2 q_o$.
Figure~\ref{f:rightmoving} shows the results obtained from a discrete eigenvalue located in each of these domains.
(The domains $\smash{D_1^-},\dots,\smash{D_4^-}$ used in Fig.~\ref{f:leftmoving} are the symmetric counterparts to $\smash{D_1^+},\dots,\smash{D_4^+}$ relative to the imaginary axis.)
The difference between $\smash{D_1^+}$ and $\smash{D_4^+}$ is that the latter collects eigenvalues close to the branch cut, 
and results in broader solitons compared to the former \cite{BLM2018}.
The difference between $\smash{D_2^+}$ and $\smash{D_3^+}$ originates from the long-time asymptotics.

Recall that the elliptic solutions of the focusing NLS equation are determined (up to translations and phase invariance) by four complex conjugate constants,
which are the branch points of the associated Riemann surface in the IST \cite{Kamchatnov}.
For the boundary conditions~\eqref{e:BCs} with $q_-=0$, two of these branch points are fixed at $\pm iq_o$, whereas the other two, 
$\alpha = \alpha_\re +  i \alpha_\im$ and its conjugate, are given by 
the system of modulation equations 
\cite{elgurevich}
\bse
\label{e:modulationsystem}
\begin{gather}
4\alpha_\re + 2(q_o^2-\alpha_\im^2)/{\alpha_\re} = V\,,
\\[0.4ex]
\big(\alpha_\re^2 + (q_o-\alpha_\im)^2\big)K(m) = (\alpha_\re^2-\alpha_\im^2+q_o^2)E(m).
\label{e:modulation2}
\end{gather}
\ese%
Here, 
$m = {4\alpha_\im q_o}/{|\alpha - q_o|^2}$, 
while $K(m)$ and $E(m)$ are the complete elliptic integrals of the first and second kind, respectively \cite{NIST}.
The trajectory described by $\alpha$ in the complex plane as $V$ varies between 0 and $4\sqrt2 q_o$ is shown by the blue dashed curve in Fig.~\ref{f:Vnzbg}
It is this curve that defines the boundary between $\smash{D_2^+}$ and $\smash{D_3^+}$.
As shown in~\cite{elgurevich,BM2016,PRE94p060201R},
the same slow modulation of the traveling wave solutions of the focusing NLS equation
also describes the nonlinear stage of MI induced by localized perturbations of a constant background.%

Importantly, each of the domains $D_1^+,\dots,D_4^+$ results in a different outcome for the nonlinear interaction.
The four cases shown in Fig.~\ref{f:rightmoving} correspond to a choice for the discrete eigenvalue $k_o$ in each of the four domains
(the precise location is identified by the red circles in Fig.~\ref{f:Vnzbg}).

In all four cases, the soliton initially travels towards the oscillatory structure with velocity $V_\zbg(k_o)$.
The simplest case is that of $k_o\in \smash{D_1^+}$ (top row of Fig.~\ref{f:rightmoving}).  
Here the soliton is transmitted through the DSW, 
and emerges as a soliton on a nonzero background.  
However, its velocity is different after the interaction, and is given by $V_\nzbg(k_o) >4\sqrt{2}q_o$. 

When $k_o \in \smash{D_2^+}$ (second row of Fig.~\ref{f:rightmoving}), we have $V_\zbg(k_o) < V_\nzbg(k_o) < 4\smash{\sqrt{2}}q_o$. 
Here the soliton is not transmitted through the DSW, and remains trapped inside the wedge. 
The velocity $V_s(k_o)$ of the trapped soliton is obtained by solving the equation $h_\im(k_o,V) = 0$, 
with $h(k,V) = h(Vt,t,k)/t$,
which has a unique root $V=V_s(k_o)$ for $k_o \in \smash{D_2^+}$ \cite{BLM2018}.

When $k_o \in \smash{D_3^+}$ (third row of Fig.~\ref{f:rightmoving}), 
we also have $V_\zbg(k_o) < V_\nzbg(k_o) < 4\sqrt{2}q_o$, and the soliton is again trapped inside the wedge. 
Here however the equation $h_\im(k_o,V) = 0$ has two solutions, for $V = V_\ast(k_o)$ and $V = V_s(k_o)$ with $V_\ast(k_o) < V_s(k_o)$
\cite{BLM2018}. 
The second of these roots corresponds to the trapped soliton, the first to the soliton-generated wake \cite{BLM2018,limitations}.

Finally, when $k_o \in \smash{D_4^+}$ (fourth row of Fig.~\ref{f:rightmoving}),  
we have $V_\zbg(k_o) < 4\sqrt{2}q_o$ but $V_\nzbg(k_o) > 4\sqrt{2}q_o$. 
Here the soliton is transmitted through the wedge and eventually re-emerges with speed $V_\nzbg(k_o)$
\cite{limitations}.
However, the equation $h_\im(k_o,V) = 0$ has a solution for $V = V_\ast(k_o)$
which gives rise to a soliton-generated wake \cite{BLM2018}.

\paragraph{Discussion}
We emphasize that there is no data fitting in Figs.~\ref{f:leftmoving}--\ref{f:rightmoving}.
Thus, the figures demonstrate excellent agreement between the numerically computed soliton velocities and those obtained from the long-time asymptotics.
In terms of the inverse problem in the IST, the properties of the soliton depend on whether the controlling phase function is the one with ZBG or that with NZBG.
Soliton trapping by a dispersive shock was also recently discussed in Refs.~\cite{prl120p144101,BLM2018},
while soliton trapping by a rarefaction wave was considered in Ref.~\cite{jmp50p091406}.%

The dynamics remain virtually unchanged if the sharp discontinuity in the IC is replaced by a smooth function which interpolates between the asymptotic values of the potential,
demonstrating the robustness of the results. 
Note also that the dynamical behavior produced by a pure step IC is markedly different in the focusing and defocusing case.
Namely, instead of the oscillatory wedge arising here, 
in the defocusing case the pure step IC~\eqref{e:stepIC} gives rise to Gibbs-like phenomena \cite{difrancomclaughlin,biondinitrogdon}.

The numerical results do not allow us to draw any conclusions regarding the 
possible presence of a soliton-generated wake in interactions between left-moving solitons and the DSW.
We were unable to observe a wake, but we cannot exclude its existence a priori.
A definitive answer can be obtained through a rigorous calculation of the long-time asymptotics, which however 
is beyond the scope of this work.

The results of this work open up a number of interesting avenues for further research.
From a theoretical point of view, an obvious open problem is the rigorous validation of the results of this work 
by explicit computation of the long-time asymptotic behavior of solutions, 
for example using the Deift-Zhou nonlinear steepest descent method for oscillatory Riemann-Hilbert problems \cite{DZ1993}.
From a practical point of view, another obvious question is whether the results of this work can be 
generalized to other NLS-type models, as was previously done in \cite{SIREV2018} for the results obtained in \cite{BM2016}.
Finally, from an even more practical point of view, an obvious challenge will be the 
experimental observation of the results of this work, perhaps in nonlinear optical fiber experimts, 
similarly to those recently conducted in \cite{arxiv2018elrandouxsuret}.

\paragraph{Acknowledgement}
We are grateful to D.\ Mantzavinos and S.\ Li for many insightful discussions.
This work was supported in part by the National Science Foundation under grant DMS-1614623.

\def\doibase{http://dx.doi.org/}
\def\reftitle#1,{\relax}


\begin{thebibliography}{12}
\bibitem{gauge}
The term $q_o^2q$ can be removed from Eq.~\eqref{e:nls} by a simple gauge transformation,
but was added so that the nonzero background solution is independent of time.
\bibitem{accuracy}
The time at which round-off error grows to $O(1)$ is about $t=15$ when working in double precision and with $q_o = O(1)$.
\bibitem{limitations}
For right-moving solitons produced by eigenvalues $k_o \in \smash{D_3^+},\smash{D_4^+}$, 
the soliton velocity (proportional to $\Re k_o$) and 
soliton width (proportional to $\Im k_o$)
forced us to choose between faster solitons that start further away from the DSW or slower solitons that start closer to the DSW, 
thus limiting our ability to observe the interations.
\end{thebibliography}
\end{document}